\documentclass[letterpaper, 10 pt, conference]{ieeeconf}  
\usepackage{cite}
\usepackage{graphicx}
\usepackage{caption}
\usepackage{booktabs}
\usepackage{float}
\usepackage{mathtools, cases}
\usepackage{amsmath,empheq}
\usepackage{eqparbox} 
\usepackage{bm} 
\usepackage{multirow}
\usepackage{lipsum} 
\usepackage{subcaption}

\IEEEoverridecommandlockouts                              
\title{\LARGE \bf
RespNet: A deep learning model for extraction of respiration from photoplethysmogram
}

\author{Vignesh Ravichandran$^{1,*}$, Balamurali Murugesan$^{1,2}$, Vaishali Balakarthikeyan$^{1}$, Sharath M Shankaranarayana$^{2}$, \\ 
Keerthi Ram$^{1}$, Preejith S.P$^{1}$, Jayaraj Joseph$^{1}$ and Mohanasankar Sivaprakasam$^{1,2}$ 
\thanks{$^{*}$ vignesh.ravi@htic.iitm.ac.in}
\thanks{$^{1}$ Healthcare Technology Innovation Centre (HTIC), IITM, India}%
\thanks{$^{2}$ Indian Institute of Technology Madras (IITM), India}%
}

\begin{document}

\maketitle
\thispagestyle{empty}
\pagestyle{empty}

\bmdefine{\bPm}{\pm}

\begin{abstract}
Respiratory ailments afflict a wide range of people and manifests itself through conditions like asthma and sleep apnea. Continuous monitoring of chronic respiratory ailments is seldom used outside the intensive care ward due to the large size and cost of the monitoring system. While Electrocardiogram (ECG) based respiration extraction is a validated approach, its adoption is limited by access to a suitable continuous ECG monitor. Recently, due to the widespread adoption of wearable smartwatches with in-built Photoplethysmogram (PPG) sensor, it is being considered as a viable candidate for continuous and unobtrusive respiration monitoring. Research in this domain, however, has been predominantly focussed on estimating respiration rate from PPG. In this work, a novel end-to-end deep learning network called RespNet is proposed to perform the task of extracting the respiration signal from a given input PPG as opposed to extracting respiration rate. The proposed network was trained and tested on two different datasets utilizing different modalities of reference respiration signal recordings. Also, the similarity and performance of the proposed network against two conventional signal processing approaches for extracting respiration signal were studied. The proposed method was tested on two independent datasets with a Mean Squared Error of 0.262 and 0.145. The cross-correlation coefficient of the respective datasets were found to be 0.933 and 0.931. The reported errors and similarity was found to be better than conventional approaches. The proposed approach would aid clinicians to provide comprehensive evaluation of sleep-related respiratory conditions and chronic respiratory ailments while being comfortable and inexpensive for the patient. 

\end{abstract}


\section{INTRODUCTION}

Measurement and monitoring respiration of an individual is useful in a plethora of clinical conditions such as pulmonary diseases and sleep-related respiratory ailments. An abnormal respiration pattern is a clinically relevant event for identifying patient deterioration under intensive care \cite{Theerawit2017}. Dysfunctional breathing refer to a group of breathing disorders that has been associated with a wide range of chronic respiratory ailments such as asthma and chronic obstructive pulmonary disease (COPD) \cite{Boulding2016}. Sleep apneas not only cause significant disruptions to sleep but also increase the risk factor for various cardiac diseases such as hypertrophy and heart failure \cite{Bradley2009}. Traditional measurement of respiration is carried out through the use of either a spirometer (airflow), pneumography (chest movement), or abdominal electromyography (diaphragm muscle activity). These measurement modalities are cumbersome and expensive for widespread adoption in a general ward or home setting. This necessitates the use of unobtrusive sensors for obtaining respiratory information from  patients in free-living conditions. Indirect measurement of respiration can be obtained from either the ECG or PPG. The amplitude, baseline and frequency modulation of the ECG and PPG signals due to respiration is well documented in literature \cite{Clifton2007}. Many consumer-grade activity monitors and smartwatches, in particular, allow for inexpensive and ambulatory monitoring of PPG. Research in the domain of respiratory information extraction from PPG has revolved around estimating respiratory rate. \par Despite the substantial diagnostic value offered by respiration rate, extracting information about the respiratory pattern from ECG and PPG would allow for a more comprehensive evaluation of sleep conditions and other chronic respiratory ailments. The accessibility and comfort of PPG make it a viable candidate for extraction of respiration information over ECG. Prinable \textit{et al.} \cite{Prinable2018} propose a novel approach of estimating tidal volume from PPG utilizing features extracted by applying various bandpass filters. However, this method was limited to using only 30 features and was evaluated on a dataset comprising of a single healthy volunteer which limits the variability for the respiration extraction task. The task of extracting respiration waveform from PPG can be formulated into a deep learning task similar to image segmentation wherein a set of filters are learned to transform an input image to a mask. U-Net is a well established deep learning network for performing image segmentation which uses a fully convolutional network utilizing “crop and concatenate” operation between pairs of encoder and decoder sections \cite{Ronneberger2015}. U-net and the plethora of networks adapted from the U-net such as ResU-Net \cite{Shankaranarayana}, V-net \cite{Milletari2016} have shown exemplary performance in the image and volumetric medical segmentation domain. Recently Stoller \textit{et al.} \cite{Stoller2018} have proposed Wave-U-Net which is used to perform sound source seperation in the 1D audio domain. This method, however, uses the rudimentary version of U-Net as proposed by Ronneberger \textit{et al.} \cite{Ronneberger2015}. A deep learning approach to extract respiratory signal from a PPG sensor would provide tremendous utility in the fitness and clinical domains. To this end, we have developed an end-to-end deep learning framework to extract the respiratory signal from an input PPG signal. This paper emphasizes a novel approach to separate a desired encoded signal contained within an input signal through training against a target reference signal.
\\
\\
\\
In summary, the contributions in this paper are as follows:
\begin{itemize}
\item {We propose a fully convolutional network to perform end-to-end respiratory signal extraction from an input PPG signal.}
\item{We study the performance of the network on two datasets using different modalities of respiratory measurement.}
\item{We extensively evaluated the proposed method on the two datasets and have compared similitude and signal reconstruction error performance against two state of the art signal processing methods.}
\end{itemize} 

\section{Methodology}\label{methodology}

\subsection{Problem formulation}
The task is to extract a respiratory signal from a given input PPG signal. The dataset $X=\{(x^{(1)},y^{(1)}),(x^{(2)},y^{(2)}),....,(x^{(m)},y^{(m)})\}$ consists of the input PPG signal $x^{(i)}$ and reference respiration signal $y^{(i)}$, where $x^{(i)} \in R^{n}$ and $y^{(i)} \in R^{n}$. The reference respiration signal $y^{(i)}$ is of the same size of the input PPG signal  $x^{(i)}$.

The proposed RespNet network is designed in the topology of a fully convolutional encoder-decoder. The encoder section take $x^{(i)}$ as input and through downsampling, it produces feature vectors $z_{1}^{(i)}$. The decoder section uses $z_{1}^{(i)}$ as its input and through upsampling produces an output predicted respiration signal $y_{pred}^{(i)}$. These are represented by equations \ref{eq:encoder} and \ref{eq:decoder}.

\begin{equation}\label{eq:encoder}
z_{1}^{(i)} = F_{1}(x^{(i)};\theta_{1})
\end{equation}
\begin{equation}\label{eq:decoder}
y_{pred}^{(i)} = F_{2}(z_{1}^{(i)};\theta_{2})
\end{equation}

where $F_{1}$, $F_{2}$ are function representations of the encoder and decoder with parameters $\theta_{1}$  and $\theta_{2}$. The decoder network outputs $y_{pred}^{(i)}$ which denotes the predicted respiratory signal. 

The parameters in the proposed architecture is optimized by minimizing the smooth $L_{1}$ loss between $y_{pred}^{(i)}$ and $y^{(i)}$. The loss function $L(X)$ is defined as: 
\begin{equation}
L(X) = \sum_{i=1}^{m}SmoothL_1(y^{(i)}-y_{pred}^{(i)})
\end{equation}

\begin{equation}
  SmoothL_1(y_{diff}) = 
  \begin{cases}
    0.5(y_{diff})^2 &\text{if} \, \lvert y_{diff} \rvert < 0 \\
     \lvert y_{diff}\rvert -0.5 &\text{otherwise,}
\end{cases}
\end{equation}



\subsection{Model Architecture}
\begin{figure*}[h]
\centering
\includegraphics[width=0.9\textwidth]{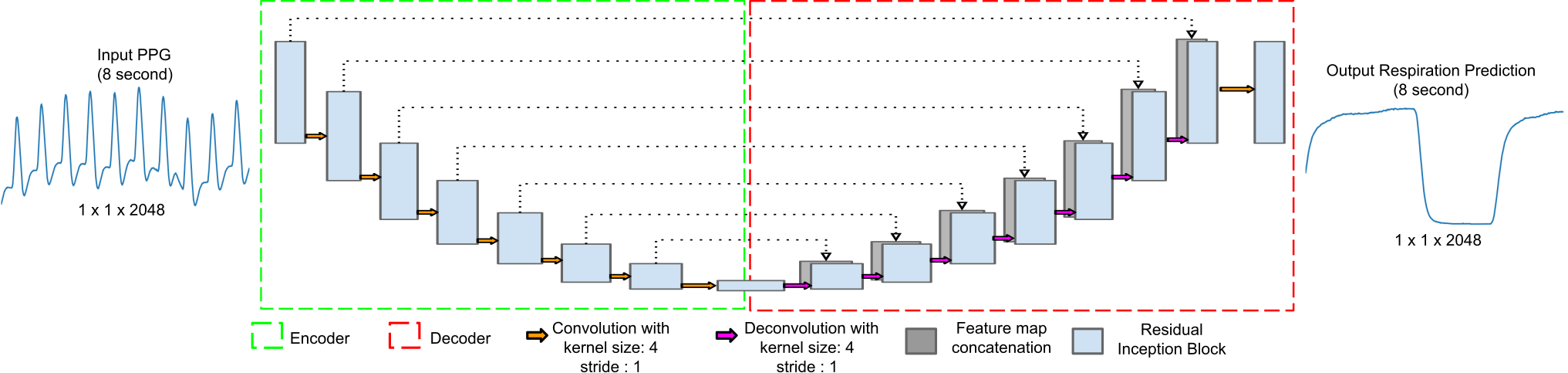}
\caption{Proposed Respiration Extraction Network: Encoder-Decoder architecture utilizing Dilated residual inception blocks}
\label{arch_respnet}
\end{figure*}

The proposed network is adapted from the IncResU-Net network \cite{ShankaranarayanaJBHI} which was made for 2D medical image segmentation application. The architecture of the proposed fully convolutional network for performing Respiration signal extraction is shown in Fig. \ref{arch_respnet}. The encoder section is divided into eight levels to perform the downsampling operation. A 1D convolution operation of size 1$ \times $4 is used to downsample the input features. Instead of carrying out the downsampling operation using max-pooling, strided convolution is used instead to improve training efficiency \cite{springenberg2015striving}. The downsampling operation decreases the input size while increasing the number of filters at each encoder by a factor of two till the number of encoder filters are 512 after which subsequent encoder levels are maintained at 512 filters. In each encoder level, 1D Convolution with stride 4 is applied followed by Batch Normalization and leaky ReLU (with slope 0.2). \par
\begin{figure}[h!]
\centering
    \includegraphics[width=0.445\textwidth]{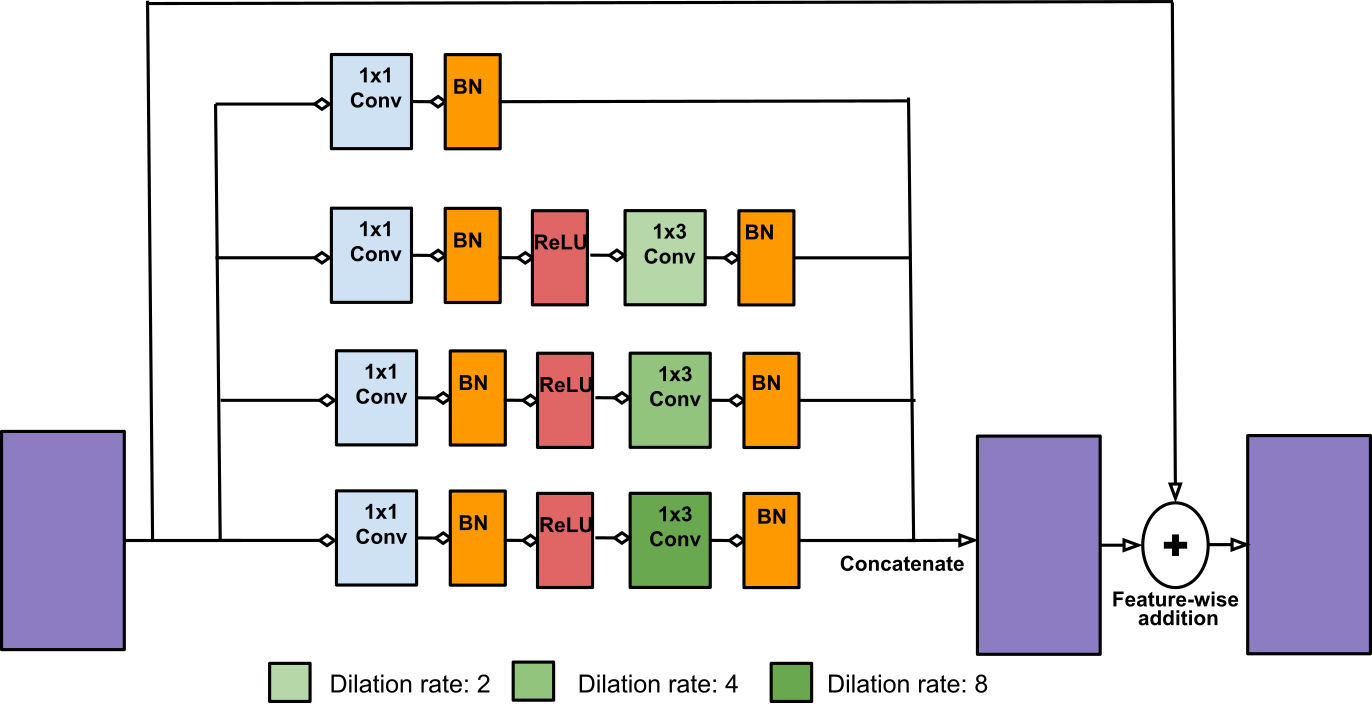}
    \caption{Dilated residual inception block}
\label{incresblock}
\end{figure}

 The output of each encoder level is then provided to the dilated residual inception block is seen in Fig. \ref{incresblock}. Usage of the dilated residual inception block provides a larger receptive field without a significant increase in parameters. Further the use of residual connections within the block is meant to greatly reduce the vanishing gradient problem and reduce the convergence time during training \cite{he2016deep}. The decoder section of the proposed network utilizes feature concatenation between the feature map of its corresponding encoder pair at the respective level similar to the original U-Net \cite{Ronneberger2015}. 
After performing convolution and dilated residual convolution operation, upsampling is performed using a deconvolution operation at each level of the decoder. In the final level of the decoder 1$\times$1 1D convolution operation is performed to map the features channels to the desired number of output channels.

\section{Dataset Description}
To perform training and evaluation of the proposed network, we make use of two distinct and publically available datasets: 
CapnoBase \cite{karlen2010capnobase} and Vortal \cite{Charlton2016}. While both datasets record ECG, PPG and a reference respiratory signal, the reference signal used in both cases are different. While the reference respiratory signal in the CapnoBase dataset was collected using capnometry, the Vortal dataset used Impedance Pneumography and Oral-Nasal pressure signals. The details about the datasets including the different sampling rate can be found in Table \ref{table_dataset1}. The CapnoBase dataset is comprised of 8 minute recordings of ECG and transmittance PPG along with capnometry from 42 subjects (13 adults, 29 children \& neonates). The data collection was performed on patients undergoing elective surgery or routine anesthesia. The Vortal dataset consists of recordings of ECG, reflectance PPG and reference respiratory signals. The dataset was collected on healthy volunteers of different age groups at supine posture. The PPG singals and reference respiration signals from both datasets were resampled to 256 Hz, this is to ensure compatibility with the proposed network which requires an input and label of size 2048 (256$\times$8). We extract 8-second length windows of PPG and reference respiration signal for both datasets and prepare an 80-20 train and test split for the respective datasets. 
Table \ref{table_dataset1} summarizes the dataset size used for training and testing.
\begin{table}[!htb]
\caption{Dataset Description}
\label{table_dataset1}
\resizebox{0.5\textwidth}{!}{%
\begin{tabular}{|c|c|c|c|c|c|}
\hline
\begin{tabular}[c]{@{}c@{}}Dataset \\ Name\end{tabular} & \begin{tabular}[c]{@{}c@{}}PPG\\ sampling rate\end{tabular} & \begin{tabular}[c]{@{}c@{}}Respiration reference\\ sampling rate\end{tabular} & \begin{tabular}[c]{@{}c@{}}Number of\\ 8-second windows\end{tabular} & \begin{tabular}[c]{@{}c@{}}Train-set\\ windows\end{tabular} & \begin{tabular}[c]{@{}c@{}}Test-set\\ windows\end{tabular} \\ \hline
CapnoBase & 100 Hz & 25 Hz & 2520 & 2016 & 504 \\ \hline
Vortal & 500 Hz & 25 Hz & 10443 & 8354 & 2089 \\ \hline
\end{tabular}
}
\end{table}

\section{Experiments and results }


\begin{figure*}[!htb]%
\centering
\begin{subfigure}{0.49\textwidth}
\includegraphics[width=\textwidth]{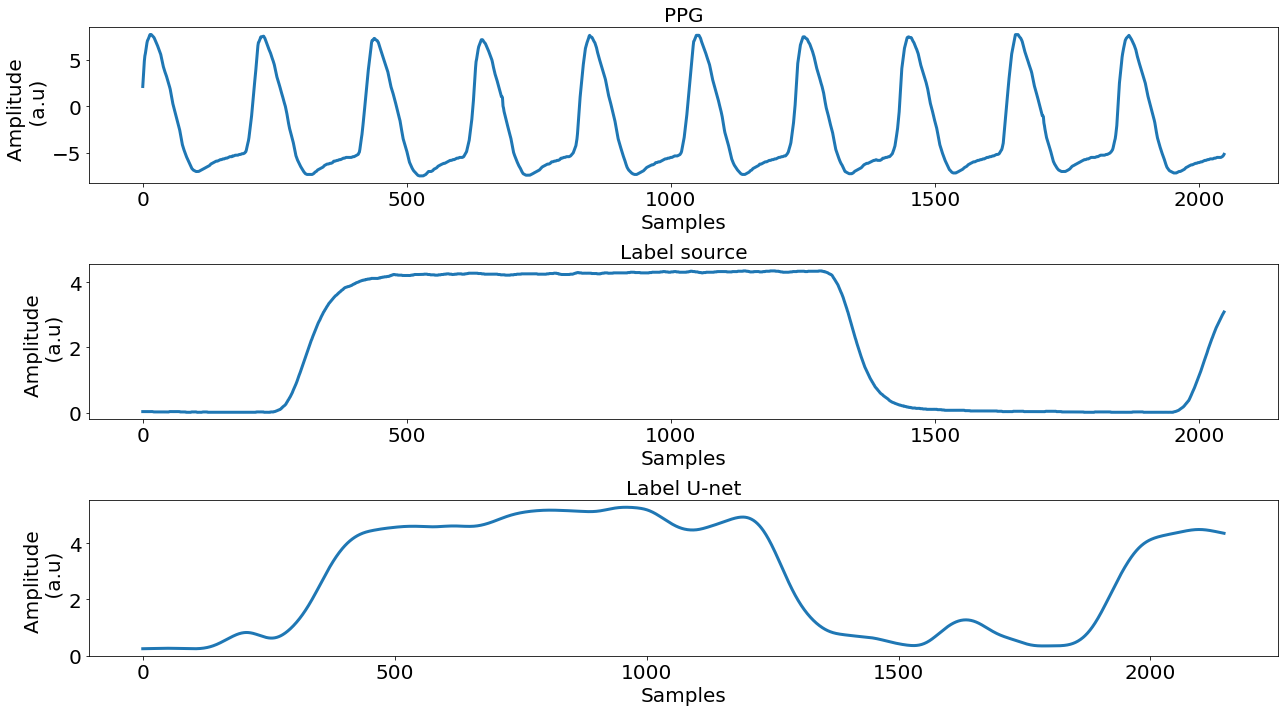}%
\caption{}
\label{fig:Ng1}%
\end{subfigure}
\begin{subfigure}{0.49\textwidth}
\includegraphics[width=\textwidth]{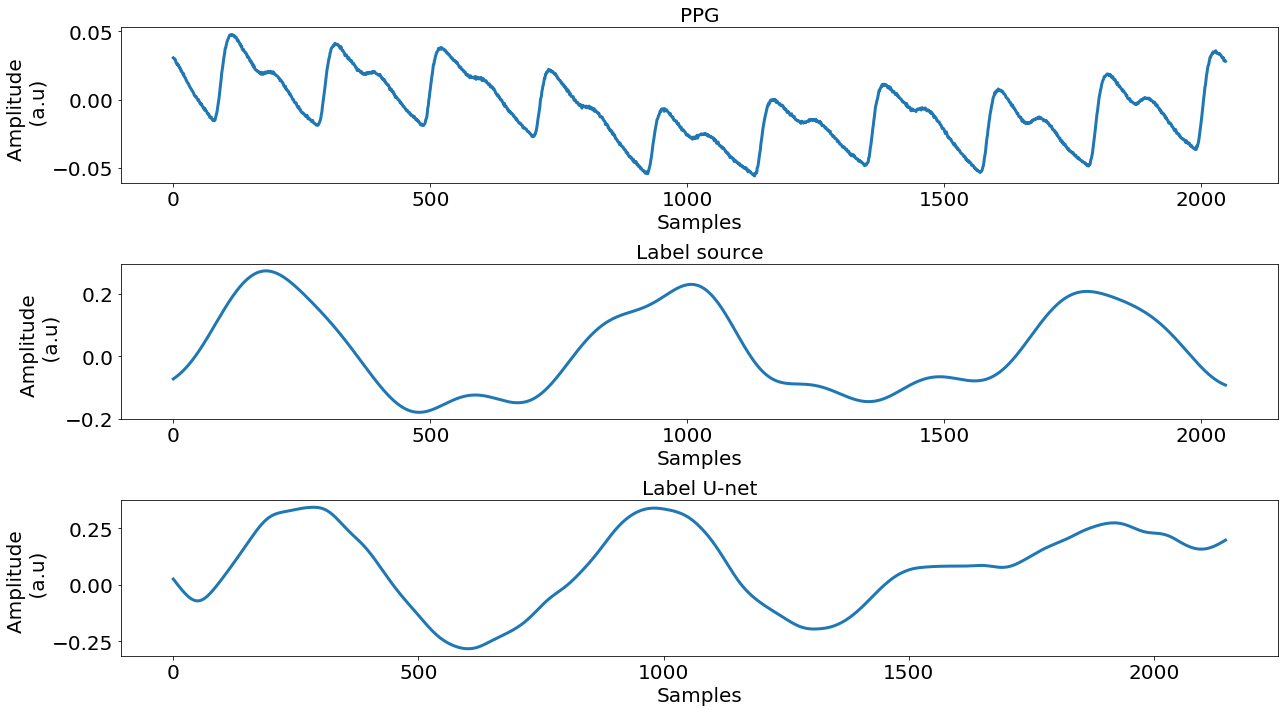}%
\caption{}
\label{fig:Ng2}%
\end{subfigure}
\caption[Two datasets evaluation]{(a) Sample PPG, reference respiration signal (label source) \& RespNet prediction (label U-net) for CapnoBase dataset \
(b) Sample PPG, reference respiration signal and RespNet prediction for Vortal dataset}
\end{figure*}

\subsection{Training Method} 
During training, the network was initialized with random weights. Smooth L1 error is determined between network prediction and ground-truth signal for each minibatch comprising of 256 input windows. The network parameters were optimized using Stochastic Gradient Descent. The learning rate of the network was set to 0.01 and the momentum to 0.7. The training was carried out for 2000 epochs. The model was developed and implemented in PyTorch \cite{paszke2017automatic}. The training was carried out in a workstation using a i7 8700K CPU and Nvidia GTX1080Ti 11GB GPU. Training and evaluation of the proposed network was carried out separately for both datasets due to the difference in sensing modality used to acquire their respective reference respiration signals.

\subsection{Evaluation Method}
The respiratory signals extracted from both CapnoBase and Vortal PPG datasets using the proposed network was validated with their corresponding ground truth respiratory signals. Further, comparative metrics were obtained through similar evaluation of amplitude modulation (WAM), frequency modulation (WFM) respiratory signals extracted from the above mentioned PPG datasets using the RRest toolbox \cite{charlton2016waveform}. Signal similarity evaluation is carried out by finding the cross-correlation and Mean Square Error (MSE), which are commonly used to measure similarity of two signals \cite{estrada2015emg} \cite{batista2001compression}. Lag is also evaluated along with cross-correlation to study if the change in respiration signal reference influences lag between the model prediction and reference. 
The WAM and WFM predictions from the RRest library is obtained using the input PPG signal. The respiration predictions provided by the library were 60 Hz, hence the respiration signal predictions obtained from the RespNet network were accordingly downsampled before evaluation. Min-Max normalization was applied to all the signals to scale the values between the range 0 and 1 before determining the comparison metrics. MSE and cross-correlation were found for the WAM, WFM and RespNet predictions against the normalized reference respiration signal for both datasets. Table \ref{results2} demonstrates the performance of the proposed RespNet network against conventional methods through mean cross-correlation, mean lags and mean MSE.

\begin{table}[]
\caption{Comparison between RespNet and other methods}
\label{results2}
\resizebox{0.5\textwidth}{!}{
\begin{tabular}{|c|c|c|c|c|}
\hline
Dataset & Method & MSE & Cross-Correlation & Lag \\ \hline
\multirow{3}{*}{CapnoBase} & WAM & 0.301 & 0.925 & 0.024 \\ \cline{2-5} 
 & WFM & 0.364 & 0.858 & 0.014 \\ \cline{2-5} 
 & RespNet (Ours) & \bf{0.262} & \bf{0.933} & \bf{0.004} \\ \hline
\multirow{3}{*}{Vortal} & WAM & 0.247 & 0.927 & 1.929 \\ \cline{2-5} 
 & WFM & 0.272 & 0.853 & 1.706 \\ \cline{2-5} 
 & RespNet (Ours) & \bf{0.145} & \bf{0.931} & \bf{0.052} \\ \hline
\end{tabular}
}
\end{table}

\section{Discussion}
As can be seen in Table \ref{results2} the proposed network, RespNet shows better performance in the task of respiration signal extraction from an input PPG when compared to conventional approaches. The respiration predictions provided by RespNet have lower MSE while having high cross-correlation and low lag with reference respiration signal for both datasets. The lower lag of the proposed network is more apparent during evaluation in the Vortal dataset wherein a different reference respiration modality (Impedance Pneumograph) was used. This also shows the unique advantage poised by such a learning method compared to standard signal processing approach which allows for adaptation towards a diverse range of respiration sensing modalities during training. Figure. \ref{fig:Ng1} shows a sample input PPG signal from the CapnoBase dataset along with the reference respiration signal and \ RespNet prediction. Figure. \ref{fig:Ng2} shows a sample input PPG signal from the Vortal dataset along with the reference respiration signal and RespNet prediction. 
\\
\\

\section{CONCLUSION}

The present work describes a novel approach to extract the respiration signal from PPG as opposed to performing respiratory rate estimation. The end-to-end deep learning framework proposed utilizes PPG signal as input and allows training with any corresponding reference respiratory signal. We report superior performance compared to traditional methods with a MSE of 0.262 and 0.145 and cross-correlation of 0.933 and 0.931 for the respective datasets. This indicates the feasibility of extracting respiratory signal from wearable devices for a variety of applications including an inexpesive approach to monitor breathing retraining exercises. However extensive training has to be carried out on a wide range of breathing anomalies and the corresponding performance study needs to be carried out. Future scope of the proposed study would be to improve network performance in detecting inspiratory and expiratory loads and exploring the feasibility of performing respiration extraction from a wrist-worn reflectance PPG sensor. Additionally, the performance of the network under mild and major motion conditions requires evaluation.





\section*{ACKNOWLEDGMENT}
The authors would like to acknowledge Dr. Peter H Charlton from King's College London for providing access to the Vortal dataset for carrying out this study and for the development of the RRest library on MATLAB.

\bibliographystyle{ieeetr}
\bibliography{main}
\end{document}